# Equilibrium Low Temperature Heat Capacity of the Spin Density Wave compound $(TMTTF)_2$ Br : effect of a Magnetic Field .

S. Sahling [a, b], J.C. Lasjaunias[a], R. Mélin[a], P. Monceau[a] and G. Rményi[a]

[a]Institut NEEL, CNRS, & Université Joseph Fourier, BP 166, F-38042 Grenoble cedex 9, France,

[b]Institut für Festkörperphysik, IFP, Technische Universität Dresden, D-01062 Dresden, Germany,

e-mail: sahling@physik.tu-dresden.de; lasjaunias@grenoble.cnrs.fr

## Abstract

We have investigated the effect of the magnetic field (B) on the very low-temperature equilibrium heat capacity $c_{eq}$ of the quasi-1 D organic compound $(TMTTF)_2Br$ , characterized by a commensurate Spin Density Wave (SDW) ground state. Below 1K, $c_{eq}$ is dominated by a Schottky-like $A_ST^{-2}$ contribution, very sensitive to the experimental time scale, a property that we have previously measured in numerous DW compounds. Under applied field (in the range 0.2- 7 *T*), the equilibrium dynamics, and hence $c_{eq}$ extracted from the time constant, increases enormously. For $B \geq$ 2-3 *T*, $c_{eq}$ varies like $B^2$ , in agreement with a magnetic Zeeman coupling. Another specific property, common to other Charge/Spin density wave (DW) compounds, is the occurrence of metastable branches in $c_{eq}$, induced at very low temperature by the field exceeding a critical value. These effects are discussed within a generalization to SDWs in a magnetic field of the available Larkin-Ovchinnikov local model of strong pinning. A limitation of the model when compared to experiments is pointed out.

## 1. Introduction

Quasi-1D conductors, among various generic properties related to their charge or spin density wave (C/SDW) low temperature ground state, exhibit specific thermodynamic properties that we have extensively investigated [1, 2, 3 ]. At very low T, i.e. below about 1K, they are determined by the presence of low-energy excitations (LEE), directly related to the inherent disordered character of the DW due to strong pinning by impurities [4, 5]. These metastable states, analyzed as two-level systems (TLS), are at the origin, in particular, of an extra- phononic contribution to the heat capacity, and slow, non-exponential relaxation with "aging" effects.

The physical interpretation of the LEEs, characterized by long time relaxation, and a broad relaxation time spectrum g(log $\tau$), was initially described by the Larkin-Ovchinnikov local model of independent strong-pinning impurities [4, 5]. Thereafter, R. Mélin et al. [6, 7] improved this model, by taking in account the interactions between the soliton excitations created at different pinning centers, which provided interpretations of specific properties like "interrupted aging", the short-time scale $T^{\alpha}$ contribution to heat capacity, the strongly time-dependent $T^{-2}$ contribution which is the consequence of the TLS energy landscape.

We have recently pointed out the strong sensitivity of the low-T heat capacity to moderate magnetic fields, in either CDW [8, 9] or SDW [10] systems. In particular, hysteretic effects can occur during the first field excursion, with B not exceeding 1*T* , as observed in two CDW compounds, namely o-$TaS_3$ and the blue bronze $Rb_{0.3}MoO_3$ [8], which results in a new metastable branch. In the case of the SDW compound $(TMTSF)_2PF_6$, with an incommensurate SDW ground state at low-T, the $c_p$ measured at 100 mK on a short time-scale, i.e. about 1 s, increases very rapidly under field, with a sharp maximum at B = 0.2 *T* ; this strong field sensitivity disappears rapidly on increasing temperature [10]. By fitting the $c_p$ data by Schottky anomalies, one can estimate the concentration of related TLS. The estimation of the equilibrium heat capacity at long time ($10^4$ s) and in zero field yields a number of TLS of 5.7 $10^{22}$ /mole, whereas at the maximum for B = 0.2 *T* , from heat release experiments, a concentration of at least 1.4 $10^{23}$ /mole is obtained [10], which means one TLS "defect" for 4 molecules!

We presently complete this study by the investigation of the *equilibrium* heat capacity $c_{eq}$ of the commensurate SDW or antiferromagnetic (AF) compound $(TMTTF)_2Br$, under magnetic field up to 7 *T*, and in a somewhat lower T-range, between 60 and 300 mK. $(TMTTF)_2Br$ is a quasi-1D conductor at room temperature, with a charge gap opening around 100 K, and which undergoes a commensurate spin density wave transition at $T_c$ = 12.1 K. In this compound, one-dimensional spin-Peierls SP) fluctuations were detected to develop below 70K and to vanish at the AF transition, indicating the competition between the AF and the SP order [12], while for other $(TMTSF)_2X$ salts with X=$PF_6$ or $AsF_6$ these fluctuations condense in the SP phase below 19K [13] revealed by satellite reflections as recently observed by neutron scattering [14]. In the AF state, $2k_F$ peak intensity of x-ray diffuse scattering has been reported as well as $4k_F$ which may result from the magnetoelestic coupling of the lattice with the AF modulation.The same sample was previously studied – but in zero-field- by short time heat capacity [15] and long heat relaxation in response to different times $t_H$ of energy delivery („waiting time technique") [3, 15, 16] .

For a better understanding of the low temperature heat capacity anomalies in 1D materials we need their exact temperature and magnetic field dependences. If the maximum of the relaxation time spectrum is longer than the characteristic time of the heat capacity measurement (as is the case in time dependent heat capacity and heat release experiments), these dependences cannot be obtained exactly, since we do not know how the relaxation time spectrum shifts due to the temperature or magnetic field influence in comparison to our experimental time scale. If for example, the spectrum shifts to longer time with increasing field, the observed heat capacity decreases, since a smaller part of low-energy excitations contributes to the heat capacity for a given measuring time, while the total heat capacity could be unchanged at the same time. In addition, from the time $t_{in}$, needed for the termination of the relaxation of low-energy excitations which then reach the quasi-equilibrium with phonons (which is close to the maximum relaxation time of the g(log $\tau$) spectrum), we can obtain detailed information on how the relaxation time depends on the temperature and on the magnetic field.

In the incommensurate materials like $(TMTSF)_2PF_6$ [1, 10] it was impossible to measure the total heat capacity since the upper limit of the relaxation time spectrum is too long. From waiting time experiments with the commensurate $(TMTTF)_2Br$ [3, 15, 16 ] we know, that the upper limit of the relaxation time spectrum is much shorter and it seems to be possible to measure the total or equilibrium heat capacity. However, even in this case the experimental requests are very hard, since we expect the equilibrium between the low-energy excitations and the phonons after the time $t_{in} = 10^3 - 10^4$ s, while the magnetic field probably increases this time. Thus, we need a high thermal resistance $R_{hl}$ between the sample and the bath of the order of $10^7$ K/W, a high resolution of temperature measurements of about 1 μK, a thermal regulation with a stability of about 1 μK, a fluctuation of the parasitic heat flow $\Delta Q_{par}$ less than 1 pW and all this during the time of $10^4$-$10^5$ s and under magnetic field. Our experiment demonstrates that these hard experimental conditions can be realized.

These new results confirm the presence of a "low-field" metastable branch, with possible hysteretic effects, and a reversible " high-field" contribution in equilibrium $c_p$ varying like $B^2$, between ≈ 2 *T* and 7 *T* . We also recover a similar huge concentration of TLS, which reaches at least about 6 "defects" per molecule, and this raises again the question of their microscopic origin. Finally, we present a discussion of available theoretical approaches.

## 1. Experimentals

Both the short-time (0.1 to 1 s) heat pulse and long-time (1s to several $10^4$ s) heat capacity measurements were performed in CRTBT (now Institute NEEL) - Grenoble on a $He^3$ /$He^4$ dilution cryostat , using the same sample equipment and experimental arrangement as it is the case of the $(TMTSF)_2PF_6$ sample, previously described in [10]. Crystals of the organic salt $(TMTTF)_2Br$ - TMTTF standing for tetramethyltetrathiafulvalene (molar mass = 601 g ) - are gathered in great number to obtain a sample of total mass of 64 mg and tightly pressed between two silicon plates (2.2 $cm^2$ in surface) by use of thin teflon ribbons. A small amount of Apiezon N grease (about 7 mg) was spread over the surface to improve the thermal contact. Contrary to the case of the $(TMTSF)_2PF_6$ salt, where the individual needles were disposed with a preferential orientation in the magnetic field, the present crystals due to their much smaller size , are deposited at random in the sample holder. The sample was mechanically connected to the thermal regulated mixing chamber by a long and thin piece of Si (4 x 0.4 x 0.02) $cm^3$. At low temperature the thermal resistance of this Si plate is so large that the total heat link between the sample and the mixing chamber is determined mainly by the electrical wire of the thermometer and heater only (superconducting NbTi wire with a copper film on the surface). The heat capacity was obtained from the temperature drift after switching on or off the heater with a constant power. After the time $t_{in}$ the relaxation of the internal degrees of freedom of the sample has reached its end and one gets the equilibrium or total heat capacity by the usual way:

$$c_p = \tau / R_{hl}, \qquad (1)$$

where τ is determined from

$$\Delta T = (|T_1 - T_0|) \exp(-t/\tau) \qquad (2)$$

with $\Delta T = T_1 - T(t)$ for the heating and $T(t) - T_0$ for the cooling process. The temperature drift was measured until the temperature was constant within the scattering of 10 μK, caused by the thermal regulation and fluctuation of the parasitic power.

The heat link $R_{hl}$ was calculated from:

$$R_{hl} = (|T_1 - T_0|) / U_h I_h, \tag{3}$$

where $U_h$ and $I_h$ are the constant voltage and current of the sample heater. Since the most interesting information is obtained from the end of the temperature relaxation, we used relative large $\Delta T$ with a maximum up to 20 % of $T_0$. Therefore, a fit function was found for the heat link data and the heat link was determined for the temperature, where $\tau$ was taken (close to the equilibrium temperature). This temperature was also used for the corresponding heat capacity.

In Fig. 1 are shown typical results for different values of the magnetic field after a constant power of the sample heater was switched off. For all curves the equilibrium temperature is 92 mK. It is seen that the time for the whole relaxation of the low energy excitations, $t_{in}$, rapidly increases with increasing the magnetic field. For 7T the relaxation of the internal degrees of freedom is finished after 6300 s and only the last 0.3 mK yields the information about the equilibrium heat capacity. Nevertheless, the very small scattering of the temperature allows us to register the relaxation of this 0.3 mK during 20000 s. The scattering of the temperature is around 10 μK, i.e. the fluctuation of the parasitic power is, in fact, less than $10^{-12}$ W (for $R_{hl} \approx 10^7$ K/W).

## 3. Results and Discussion

### 3.1 Short time heat capacity

The first step in our measurements was the comparison of the *short-time* heat capacity, in zero field, obtained by the heat pulse technique, on a time span of a few seconds, with the previous data obtained on a different cryostat and sample arrangement (different heat link) presented in [2, 3, 15: and here referred as to "2s-1995"] together with present data (see Fig. 2). Short time $c_p$ is analyzed, as usually for these quasi-1D systems, by the sum of three contributions:

$$c_p = c_D + c_{ql} + c_S \tag{4}$$

where $c_D$ is the phonon contribution

$$c_D = \beta\, (T/K)^3, \tag{5}$$

$c_{ql}$ is a "quasi-linear" power law contribution in the form of

$$c_{ql} = b\, (T/K)^{\nu}, \tag{6}$$

with $\nu < 1$

and $c_S$ is a Schottky contribution

$$c_S = A_S\, f_s(x) \tag{7}$$

with

$$f_s(x) = x^2 \exp(x) / (1 + \exp(x))^2, \tag{8}$$

$$x = E_s/k_B T, \tag{9}$$

and $A_S = N_S\, k_B$, (10)

where $N_S$ is the number of two-level systems (TLS) and β, b and $N_S$ are constants. Thus, there are 5 fit parameters. Note that in the case of x << 1, the Schottky contribution $c_S$ (T) reduces to a $T^{-2}$ variation (relation 8).

Fitting yields (closed circles for present data) :

$$c_p = 8\,(T/K)^3 + 2.7\,(T/K)^{0.44} + 0.014\,(T/K)^{-2} \quad (mJ/\,mol\,K) \tag{11}$$

which is in close agreement with earlier data [15]: by about 3 % for b and β , and same value of ν; the $A_S$ amplitude is larger in present " 2005" data, but this term is very sensitive to the time span of experiments, and the slightly larger amplitude (0.014 instead of 0.011) is probably due to the larger present time span for pulse experiments in this T-range: around 3-4 s for "2005", instead of around 1s for "1995".

The short time heat capacity is strongly dependent on the magnetic field. Fig. 3 shows the dependence of the heat capacity on the magnetic field at 80 mK. As already indicated in Introduction, the high sensitivity to the magnetic field at low temperature seems to be a characteristic feature of 1D materials with SDW or CDW, since it was observed in all previously investigated materials [8, 9, 10]. It was also observed in unpublished results of the spin-Peierls system $(TMTTF)_2PF_6$. Fig. 3 demonstrates also the strong time dependence of the heat capacity. An increase of the time span from 2 to 50 s increases the heat capacity remarkably. A common property of other DW systems is the minimum of the short time $c_p$ at around 0.5 ÷ 1 *T* [8, 9, 10]. As seen in Fig. 3, this minimum disappears on increasing the time scale. At equilibrium, two behaviors occur: a constant $c_0$ value at low field, and a quadratic $B^2$ regime above 2 *T*.

## 3.2 Equilibrium heat capacity

### 3.2.1 Schottky contribution

The equilibrium heat capacity in zero magnetic field is presented in Fig. 2 and allows us to obtain the exact temperature dependence of the heat capacity. Surprisingly, we can fit the data without any quasi-linear term in Eq. 4:

$$c_p = 8\,(T/K)^3 + 1.0\,(T/K)^{-2} \quad (mJ/\,mol\,K). \tag{12}$$

Note, that the amplitude of the Schottky term of the heat capacity is, at the equilibrium, 71 times larger than the corresponding term measured from short time (2 s) data.

The heat capacity as a function of temperature below 300 mK for different magnetic field is shown in Fig. 4. In this temperature range the Schottky term of the equilibrium heat capacity dominates and shifts as expected to higher temperature with increasing magnetic field. However, the maximum of the Schottky term always occurs at too low temperature to be observed in our experiment, and all data can be fitted by $T^{-2}$ laws. In the absence of observation of the complete Schottky contribution, it is difficult to estimate the splitting $E_S$, or the concentration $N_S$ of the TLS. In a theoretical attempt for the interpretation [17], the model suggests a constant $N_S$ for the field range where $c_p$ varies like $B^2$ (Zeeman field coupling model). In order to estimate $N_S$ , we have fitted the data for B = 4 *T* , an intermediate field value in this range, by a simple Schottky anomaly, with a unique generacy $g_1/g_0$ =1, with the highest possible level splitting $E_S$ , corresponding to the departure from $T^{-2}$ at the highest possible temperature (see Fig. 4). Such a fit yields $E_S/k_B$ = 52 mK, with a minimum value $A_S$ = 50.5 J/mol K. This amplitude largely exceeds the limit value of R = $Nk_B$ = 8.32 J/mol K expected for one TLS "defect" per unit cell and yields roughly 6 TLS per unit cell. A rather similar result was obtained from long time heat release data of $(TMTSF)_2PF_6$, where the number of TLS was one for 4 unit cells [10].

To proof the field dependence of the Schottky contribution according to Eq. 8 we determine the temperature T*, where the heat capacity reaches the same value $c_p(T^*, B)$ = 1 mJ/mol K. In the case of a Schottky term we expect from Eqs. 7 - 10:

$$T^* =(A_s/c_s)^{0.5}(E_s/k_B), \qquad (13)$$

In fact, we see from Fig. 5 that the experimental data yield a linear dependence of T* for B larger than 1 or 2 *T*, i.e. the Zeeman splitting is proportional to the magnetic field. At smaller field, T* is influenced by some zero field splitting, since a Schottky contribution was observed in zero field too.

### 3.2.2 Metastable branches and Zeeman field splitting.

The zero field heat capacity depends on history. For demonstration we show in Fig. 6 the equilibrium heat capacity at 92 mK. After the first cooling without magnetic field the heat capacity in zero field is 120 mJ/mol K. A first field treatment up to 4 *T* does not change this value (see Fig. 6, branch 1, black full squares). But, after a further application of field, firstly at 4 *T* for 2 days, and then at 5.5 *T* for one day, with a temperature maintained between 70 and 90 mK, this changes the zero field value to 404 mJ/mol K. This new "metastable branch" (curve 2 in Fig. 6) was very robust against further high field excursions up to 7 *T* (red circles), the temperature being maintained below 250 mK. We see that the difference between the upper and lower branches disappears for B ≥ 3*T*. We succeeded to "anneal" or revirginize, the system by a rather non-trivial way. First, we reheated the sample under zero field at T ≈ 20 K, i.e. well above the SDW transition (at around 12 K), but we still recovered the upper branch on cooling again. Then, we have measured different points step by step, at T = 92 mK, increasing B up to 1.5 *T* (blue triangles), always in the upper phase; on decreasing B to 0.2 *T*, finally the initial lower branch was recovered (the blue triangles of the lower branch); the 1 *T* and 1.5 *T* data were obtained in these conditions.

In the case of o-$TaS_3$, a CDW system with the Peierls transition at $T_P$ = 210 K, a similar new "metastable branch" occurs under application of a field of about only 0.3 *T*, in the same T range of 100 mK. Also, this branch is very robust: it was necessary to reheat the sample above $T_P$ to recover the initial lower branch [8]. In the present case, the "annealing" temperature is ten times smaller, and this heat treatment appears insufficient to recover the initial state, without a further field application.

We have also reported in Fig. 6 different attempts to analyze the data. The lines through experimental data correspond to a very simple analysis, as the sum of two (independent) contributions:

$$c_p= c_0 + \alpha B^2 \qquad (14)$$

with $c_0$ independent of the field [$c_0$ = 120 or 404 mJ/ mol K, for the lower and the upper branches, respectively], and a common quadratic $B^2$ contribution, with $\alpha$ = 190 $(B/T)^2$. The change in the value of $c_0$ characterizes the metastability, while the quadratic term is independent of this metastability.

A third attempt given by line 3 for the upper branch, is obviously non adequate, but interesting for comparison. It corresponds to the assumption of an energy splitting of the form: $E_S = E_0 + \gamma B$, which is introduced in the Schottky formalism (relations 5, 6, 7). $E_0$ is the zero-field splitting, in addition to a field induced Zeeman splitting proportional to B. For the high-T tail of the Schottky : $c_S \propto x^2$ , or $\propto (E_0 + \gamma B)^2/T^2$ . The disagreement with the experimental data shows that this ansatz does not work, i.e. the zero field Schottky contribution cannot be explained by a zero field splitting of the same TLS which causes the Schottky term in field.

### 3.2.3 Dynamical properties

We have shown that in $(TMTTF)_2Br$ low-energy excitations exist with a broad spectrum of relaxation time up to very long values [15, 16]. We will consider now the following question: what is the process of the unusual long relaxation time, is it a tunneling or a thermal activation process? We have verified that the achievement of relaxation of the internal degrees of freedom, $t_{in}$, is proportional to the maximum relaxation time of this spectrum (peak in the g(log $\tau$) spectrum). The data for B = 7*T* in Fig. 1 can be analyzed including a long time exponential decay ($\tau_{max}$) related to $t_{in}$ as $t_{in} = 4.6\ \tau_{max}$. Such a relation has also been obtained for other temperature and magnetic field. Thus from $t_{in}$ we can find the temperature dependence of a fixed relaxation time. Fig. 7 demonstrates that $t_{in}$ (and consequently $\tau_{max}$) follows the Arrhenius law, i.e.:

$$t_{in} = \tau_0 \exp(E_a/k_B T). \qquad (15)$$

This is very different from structural glasses, where the long relaxation time of TLS at temperatures below 1 K are caused by tunneling [18]. The very low activation energy $E_a/k_B$ = 0.5 K is nearly the same for the two different branches and magnetic field independent. (see fig. 7). The activation energy is in an excellent agreement with the result of "waiting time experiments" [15, 16]. In this investigation the relaxation time spectrum was obtained, which was fitted by two Gaussian distribution centered on the two characteristic values $\tau_{fast}$ and $\tau_{slow}$. Their temperature dependences follows also Eq. (15) with exactly the same $E_a/k_B$ = 0.5 K.

Due to the strong temperature dependence, $t_{in}$ increases very rapidly with decreasing temperature. For example, $t_{in}$ reaches $10^{11}$ s at the temperature of the expected maximum in the Schottky curve, shown in Fig. 4 for 4*T*, i.e., even if we could reach a sample temperature of 22 mK, we could not determine the equilibrium heat capacity due to the very long $t_{in}$. With increasing magnetic field the maximum shifts to higher temperatures. However, at the same time, $t_{in}$ increases with the magnetic field too (see fig. 7 and 8). Probably there is no possibility to determine the maximum of the Schottky contribution, since $t_{in}$ remains always too long. Surprisingly the increase of $t_{in}$ with magnetic field is not caused by a change of $E_a$ but by a strong field dependence of $\tau_0$ proportional to $B^2$ for B > 2 *T* (see Fig. 9).

## 4. Discussion

Recently, a model of SDW phase defects generated under magnetic field was shown to explain the main properties described in this study. This model is within the continuation of the theoretical background initiated by A.I. Larkin, Yu. Ovchinnikov et al. [4, 5], who ascribed the extra-phononic excitations and their slow dynamics in SDW/CDW to low-energy excitations, modeled by TLS, originating at the strong pinning centers of the DW.

The present model [17] is based on interacting bi-solitons (pairs of solitons-antisolitons, see also ref. [7] ) generated by strong pinning impurities. It was already successful for the interpretation of several properties of zero-field experiments that we recover in this study such as (i) the $T^{-2}$ dependence (or Schottky specific heat) of the low-T specific heat, and its high sensitivity to the experimental time span as $t_H$, (ii) the thermally activated behavior of the dynamics, in agreement with this "classical" framework. Indeed, the transfer of energy between the two states occurs by classical activation over the barrier. The energy landscape is a distribution of energy splitting and barrier heights. (iii) the existence of an upper bound $\tau_{max}$ in the relaxation time spectrum - this is equivalent to the present $t_{in}$ , or $\tau_{slow}$ in previous descriptions [16]. As pointed out by Ovchinnikov et al. [5], artificially large impurity concentrations are required in the local model of strong pinning in order to fit the very low temperature specific heat experiments. As we discuss below, our extension of the Larkin-Ovchinnikov model to SDWs in a magnetic field suffers from the same weak point. Interestingly, it was pointed out by Artemenko and Remizov [19] that weak impurity potentials at high energy amount to cutting the chains into Luttinger liquids of finite length at very low temperature [20], resulting in characteristic features in the dependence of the specific heat on magnetic field [19]. Impurity effects are thus expected to be enhanced by Coulomb interactions, which might be the clue to the discrepancy in the comparison of the Larkin-Ovchinnikov model to experiments on the evaluation of the concentration of defects. We discuss here experiments on the basis of the Larkin-Ovchinnikov model without Coulomb interactions, while Artemenko and Remizov [19] discussed related experiments with Coulomb interactions without CDW or SDW instability. Incorporating rigorously Coulomb interactions in the Larkin-Ovchinnikov model beyond qualitative arguments is an open theoretical question that we do not address here.

In the extension of this classical pinning model to the case of applied field without Coulomb interactions, we proposed a mechanism of deconfinement of the bi-solitons, initially located at the pinning impurities, which results in the creation of new solitons under field. A soliton in a SDW carries a magnetic moment (see Fig. 10a) and acquires a Zeemann energy in a magnetic field (see Fig. 10b). The initial solitons (of the Larkin-Ovchinnikov model) do not bear a magnetic moment, whereas the new generated solitons, above a critical field value $B_c$ , bear a net magnetic moment $\mu$ which couple to the applied field in a Zeeman field coupling, resulting in energy levels $\Delta = g\ \mu_B$ B. This results in the characteristic $B^2$ behavior of $c_p$ at large field, exactly for B > $B_c$ . By increasing magnetic field, the Schottky anomalies shift linearly towards higher T (this property is verified by the linear variation of $T^*$ with B in Fig. 8, for B > 2 *T*), with an amplitude $A_S$ which remains constant. A minimum value for $A_S$ is estimated from a fit of the series of data of Fig. 6 above 2 *T* with a constant amplitude of $A_{Smin}$ ≈ 60 J/mol K, so even larger than our rough estimation (≈ 50 J/mol K) for the 4 *T* data ), which yields about 6 "defects" per molecule or unit cell. The defects can be explained qualitatively by distorsions in the phase configuration upon application of a field larger than $B_c$, and, at the same time, magnetic moments are carried by the random glassy-like phase pattern (see Fig. 11). The large number of defects obtained experimentally shows that the Larkin-Ovchinnikov model underestimates the effects of disorder, which may be due to interactions, as mentioned above, or to the approximation by which the energy landscape is reduced to two level systems (multilevel systems are also possible, but they are usually approximated as two-level systems as in Ref. 5).

An important property explained by the model is the change of regime of $c_{eq}(B)$ in increasing field from a constant value to the $B^2$ dependence; this occurs at the cross-over field $B_c$ where the spin-polarized solitons appear, in addition to the initial spinless ones. Above $B_c$, calculated as $2 \div 4$ *T* , in agreement with the experiments, the $B^2$ contribution is due to these spin-up polarized solitons. However, we have no explanation within the model, why $\tau_0$ in Eq. 11 and as consequently the relaxation time (including $\tau_{max}$ and $t_{in}$) depends strongly on B (proportional to $B^2$ for B > 2 *T*), while the activation energy $E_a$ is nearly field independent.

A final property is the appearance of a new metastable branch under field, at the origin of hysteretic effects. Above $B_c$ , the proliferation of solitons generates a “SDW glass” with a random phase configuration, characterized by phase defects which can be viewed as domain walls; these new defects are very difficult to anneal. When the field is reduced back to zero, there remains an additional contribution, corresponding to a new value of the constant $C_0$ , and possible hysteretic effects.

**Conclusions:**

Like in other 1D materials with CDW or SDW the heat capacity of $(TMTTF)_2Br$ is below 1K strongly time dependent. However, by contrast to all other investigated systems, we could find in our very sensitive and long time heat capacity measurements of $(TMTTF)_2Br$ the end of the relaxation of the low-energy excitations and determine for the first time the total equilibrium heat capacity and its dependence on temperature and magnetic field.

The equilibrium heat capacity can be described by the $T^3$ term of phonons and a $T^{-2}$ term, without any quasi-linear term necessary to fit the time dependent heat capacity at short time. The $T^{-2}$ term shifts with increasing magnetic field to higher temperature as expected for a Schottky anomaly. The data allow to estimate the number of TLS, which is very high, of the order of 6 TLS per unit cell. The large number of TLS can be explained by a large number of deconfined bi-solitons, which largely exceeds the number of local impurities [17]. However, the Larkin-Ovchinnikov approach and its generalizations require too strong concentrations of defects to match experiments. The interpration of the amplitude of the total entropy in the Schottky anomaly as measured in the present experiment (disorder versus electronic spin) remains an unsolved question which needs further investigations

The heat capacity in zero magnetic field depends on history. The heat capacity measured after a first zero field cooling is roughly 4 times smaller than the heat capacity measured after a field treatment with B > 5 *T*. This field induced branch is very stable. To come back to the lower branch, the sample must be heated at least to a temperature above 20 K.

The end of the relaxation of the low energy excitations, $t_{in}$, is proportional to the upper limit of the relaxation time spectrum and follows the Arrhenius law, i.e., the long relaxation time of the low-energy excitations in 1D materials is caused by a thermal activated process. This is a remarkable difference to structural glasses, where at low temperatures the tunneling process dominates [18]. The activation energy $E_a/k_B = 0.5$ K can be obtained from the Arrhenius law. This activation energy is nearly magnetic field independent, while $t_{in}$ is proportional to $B^2$.

**Figures :**

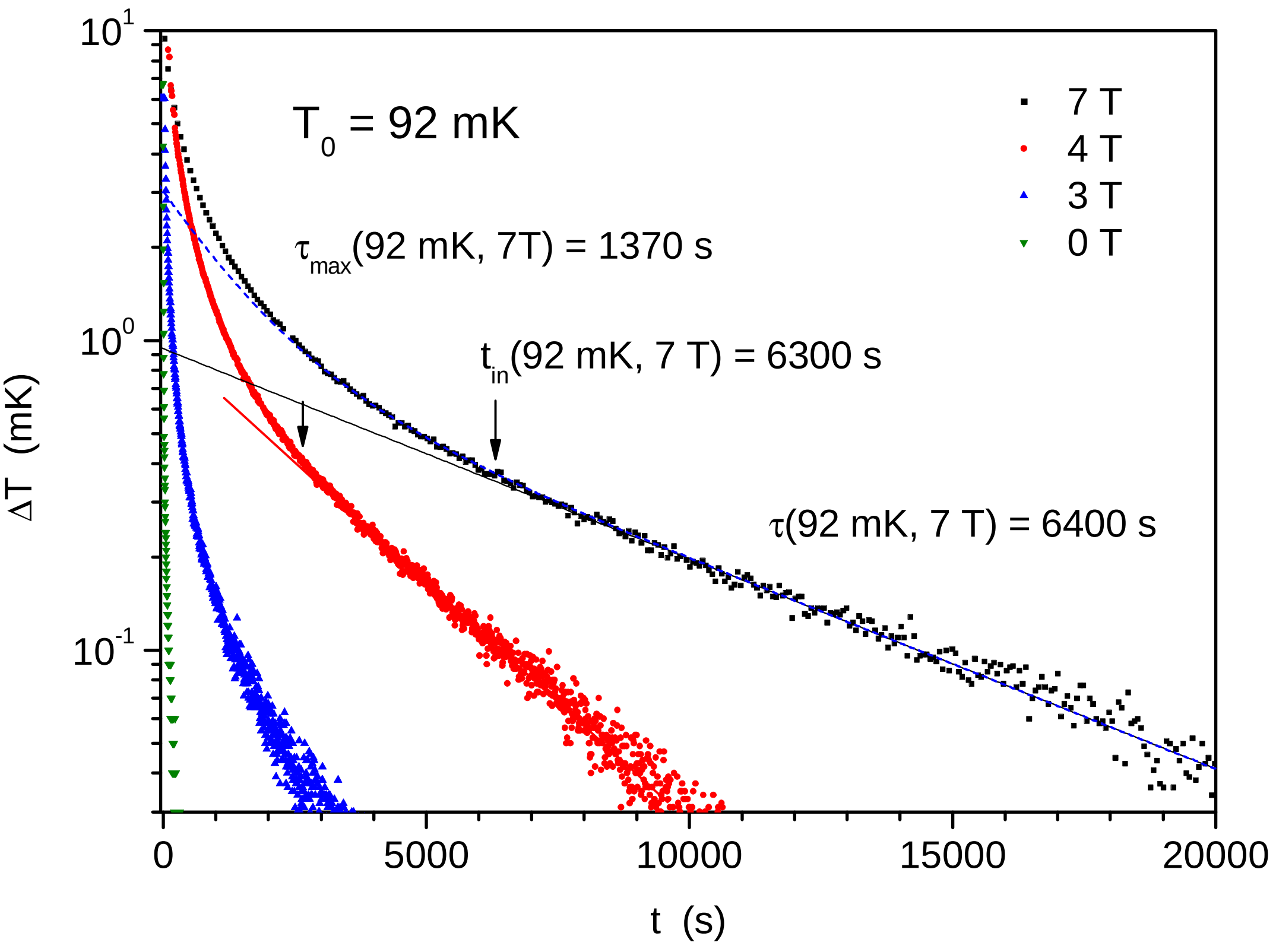

**Fig. 1:** The relaxation of the temperature from the equilibrium temperature $T_1$ = 102 mK to the new equilibrium temperature $T_0$ = 92 mK after a constant heat power to the sample was switched off for different magnetic fields. The arrows indicate $t_{in}$, the end of the relaxation of the low energy excitations, or two-level systems (TLS) . For $t > t_{in}$ the phonons and TLS are in a thermal equilibrium and relax together with the relaxation time $\tau_{eq}$ , determined by the heat link $R_{hl}$ and the heat capacity of the sample. $\tau_{eq}$ increases dramatically from 200s in zero-field to 6400s under 7*T* . Note that for this field, the relaxation was recorded over 2.2 $10^4$ s to obtain an accurate determination of $\tau_{eq.}$ The dashed line for the 7*T* shows the TLS relaxation calculated with the maximum relaxation time (peak in the g(log τ) : $\tau_{max}$ = 1370s of the low temperature excitations at 92 mK.

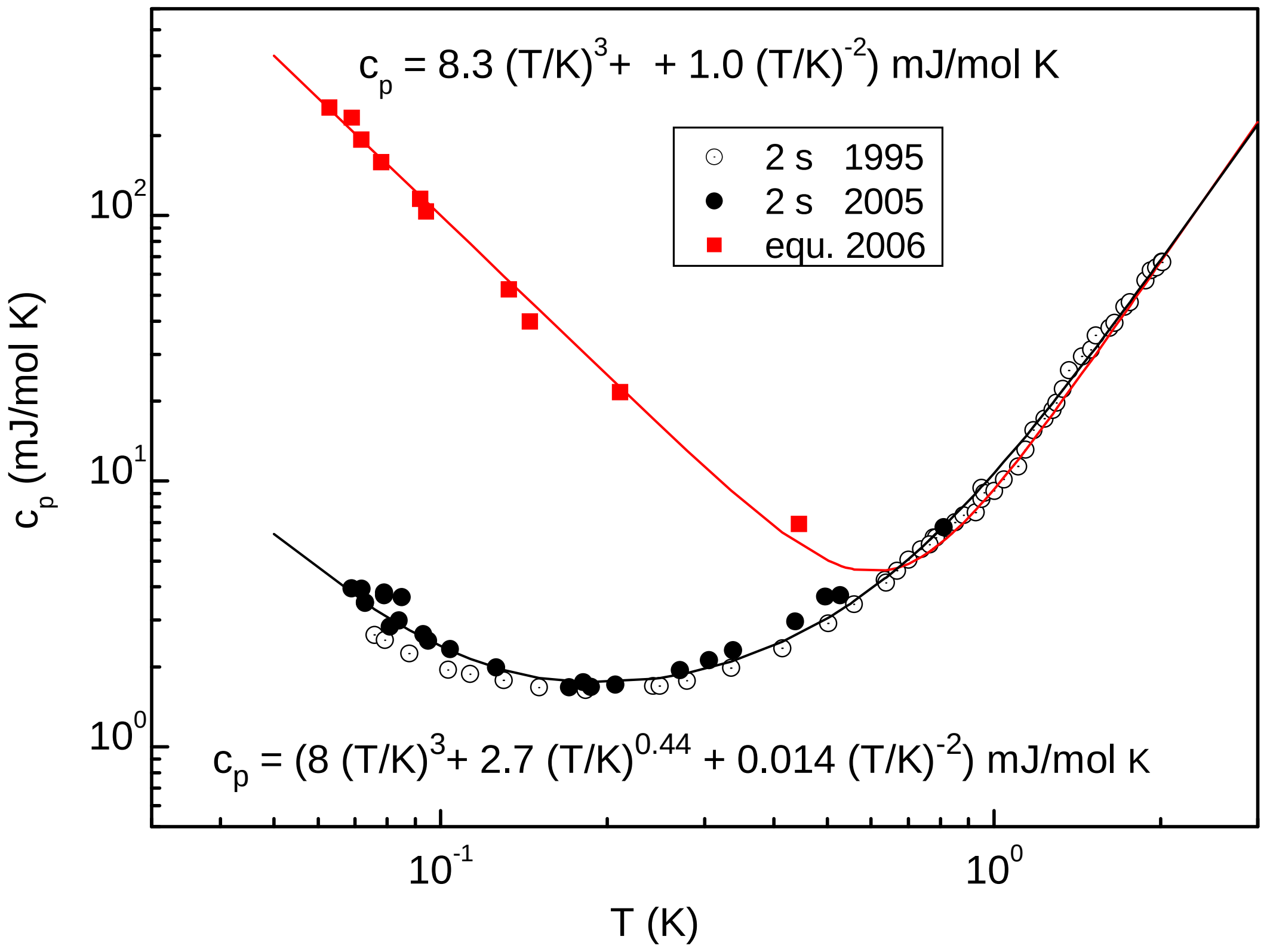

**Fig. 2:** Heat capacity as a function of temperature in zero-field: comparison between the heat capacity data of $(TMTTF)_2Br$ obtained either on the long-time scale (equilibrium) or short time (2s) scale (heat-pulses technique). The large effect of time scale appears below 1K, where the $T^{-2}$ contribution becomes predominant over the phonon one. For short-time scale data, an excellent agreement is found between the two series of data " 2s-1995" (ref.15) and present ones " 2s-2005", performed on different cryostats, thermal heat links and thermometry. The equilibrium heat capacity $c_{eq}$ is well fitted (continuous line) by only two contributions: phonons and $T^{-2}$ (tail of a Schottky anomaly), whereas for short-time, one needs a third contribution as a power law $T^{0.44}$ (see text and ref.15).

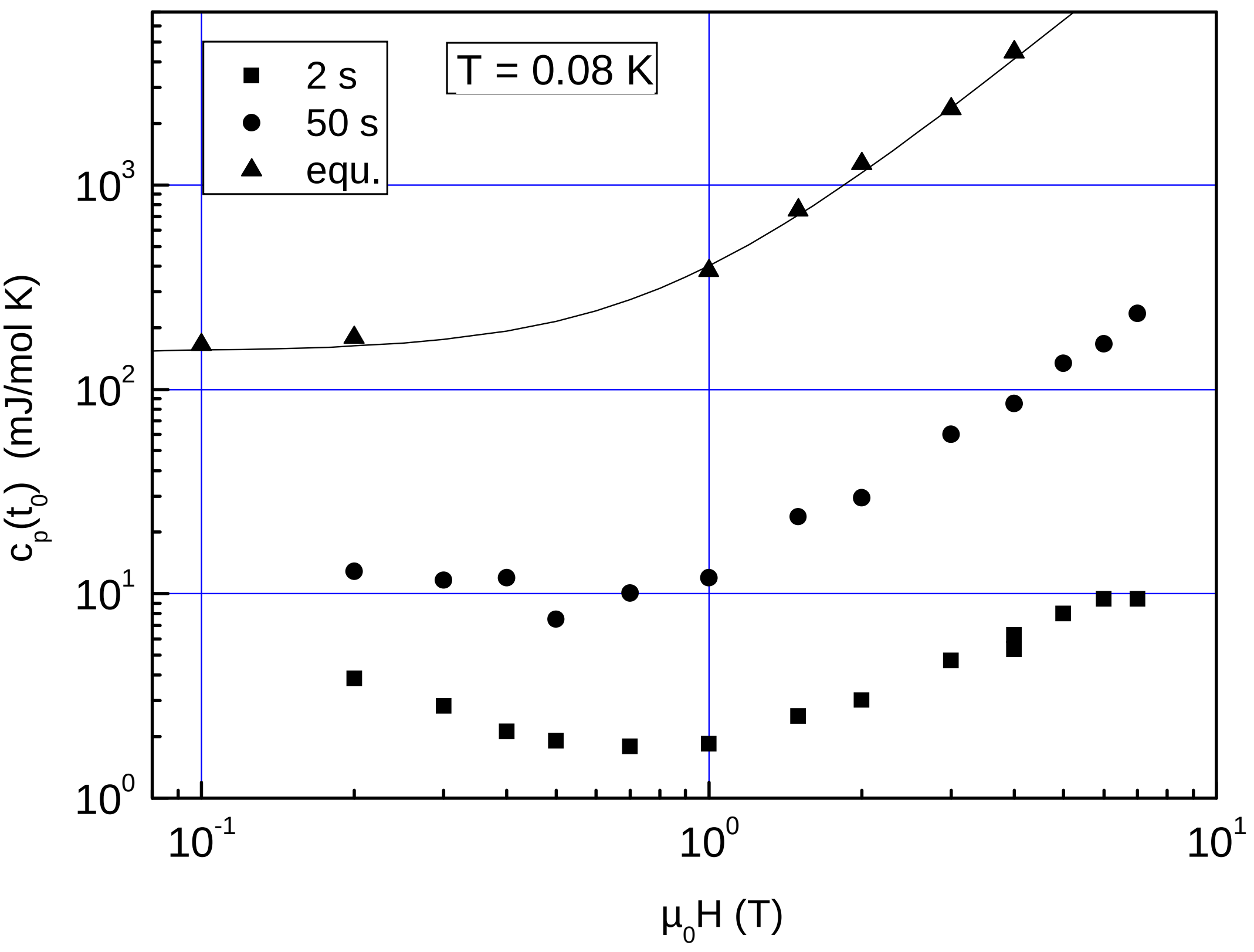

**Fig 3:** Effect of the magnetic field on the heat capacity of $(TMTTF)_2Br$ at T = 80 mK for three different durations of heat delivery $t_H$ : from short time "2s" corresponding to $t_H$ (heat pulses) of about 20 ms , intermediate 50 s, and long-time equilibrium. For field above 1*T* , $c_p$ systematically increases with B, reaching a $B^2$ regime for $c_{eq}$ . At shorter time, a minimum occurs around or below 1*T*. We note that due to the cumulated effects of both $t_H$ and B, $c_p$ varies by 3 orders of magnitude for a field around 5 *T*

.

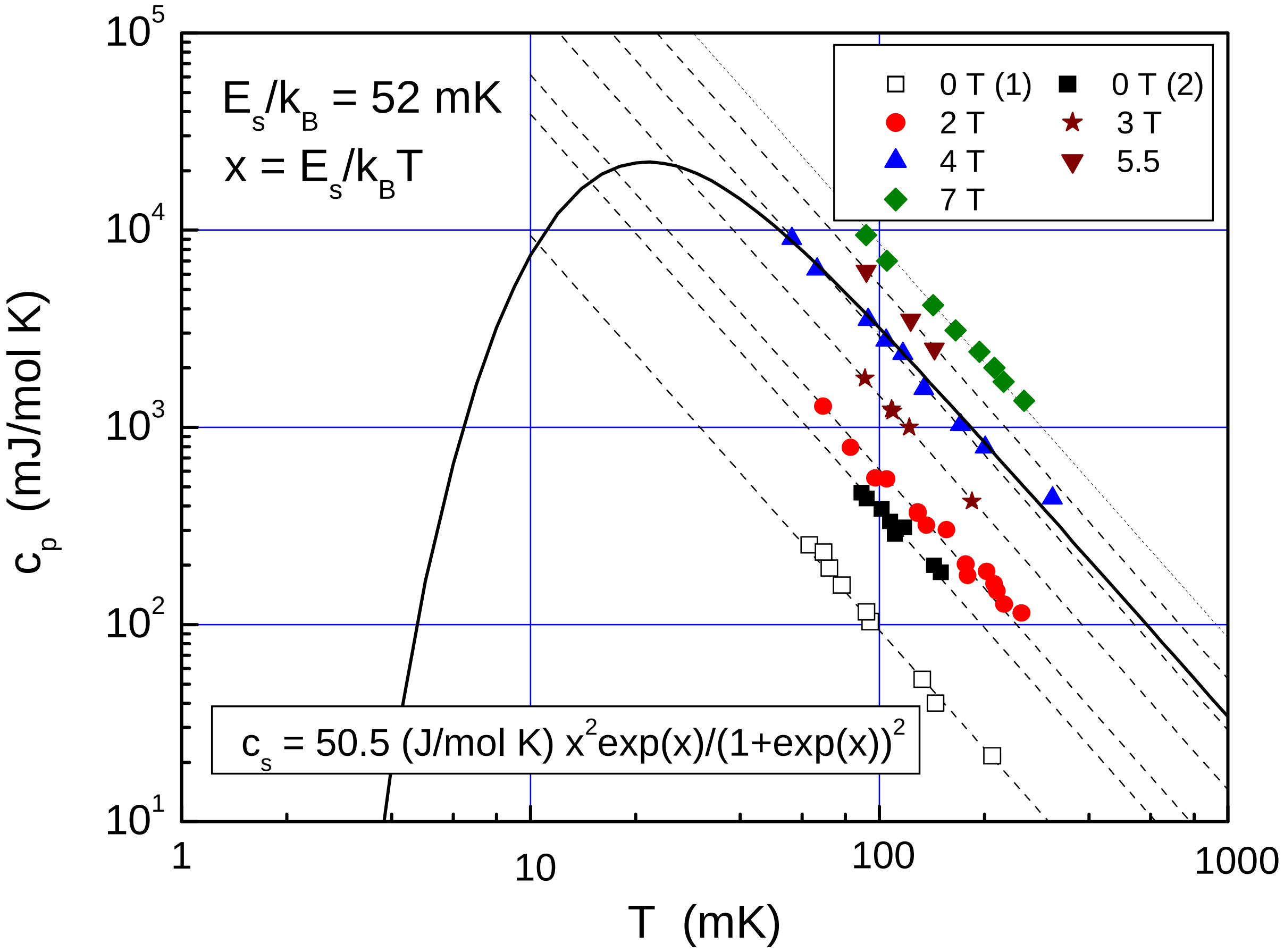

**Fig 4:** The heat capacity as function of temperature for different magnetic fields. All data , for every B value , can be fitted in this T-range by $T^{-2}$ laws (straight lines) representing the high-T tail of the Schottky specific heat. For zero-field, the two symbols correspond to the two different branches, as explained in the text. For B = 4*T*, a tentative fit by the usual Schottky anomaly (with degeneracy $g_1/g_0$ = 1) was done corresponding to the minimum possible amplitude. It yields a two-level state splitting $E_S/k_B$ = 52 mK, and a number of states per mole (per unit cell) of $(TMTTF)_2$ Br of about 6.

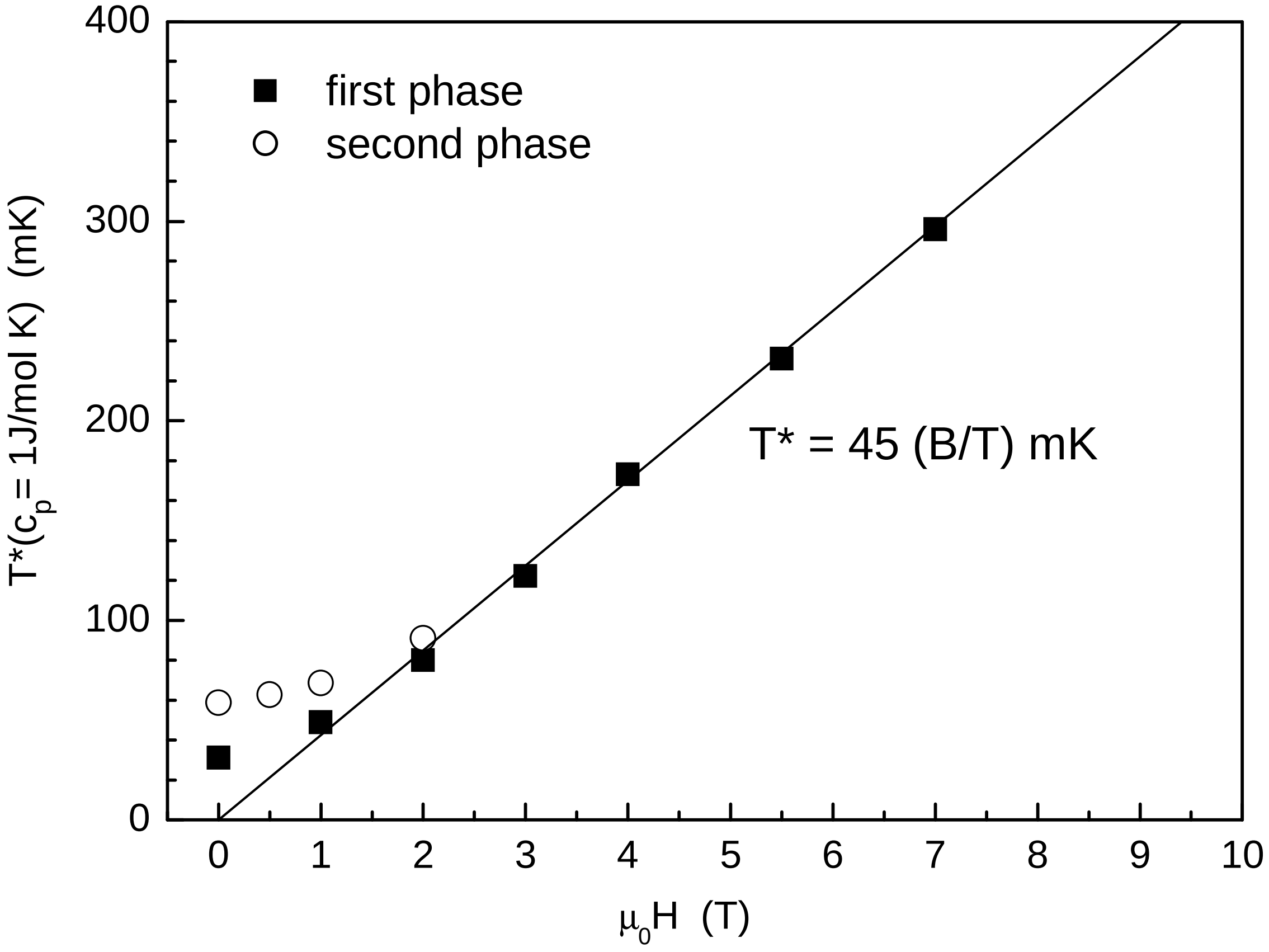

**Fig. 5:** Variation of the temperature at different fields; $T^*$ is defined for a same value of $c_p$ (here 1 J/mol K). The linear variation for B > 2*T* ( $T^* = 45$ ( B/*T*) mK) indicates a Zeeman field splitting, in agreement with the quadratic $B^2$ regime of $c_p$. For B < 2*T* , the departure from linearity indicates a different physical origin. The low-field range corresponds to two different phases (or $c_p$ branches).

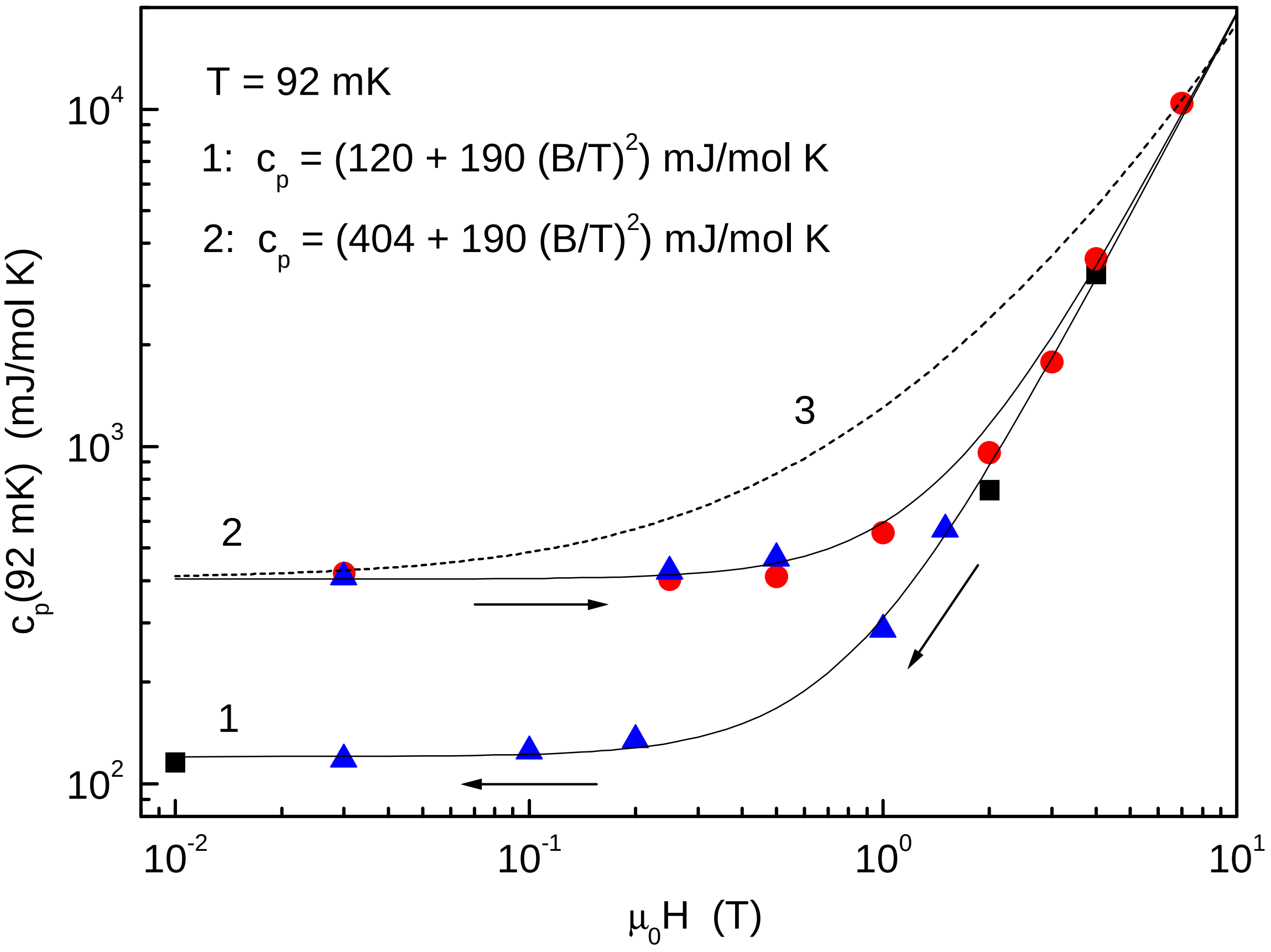

**Fig. 6:** Equilibrium heat capacity at T = 92 mK as a function of magnetic field, with two metastable branches for field below 2*T* . The different symbols refer to different experimental histories (see text for details). The metastability of the low-field data is shown: starting from the initial lower branch 1, one recovers the upper branch 2 after exposure to high field ( > 5*T*) at low T ( < 100 mK). For B above 2*T*, $c_p$ obeys the same quadratic $B^2$ variation. Solid lines 1 and 2 through experimental data are fits by $c_p = c_0 + \alpha B^2$ . Relation for dashed line 3 is explained in the text.

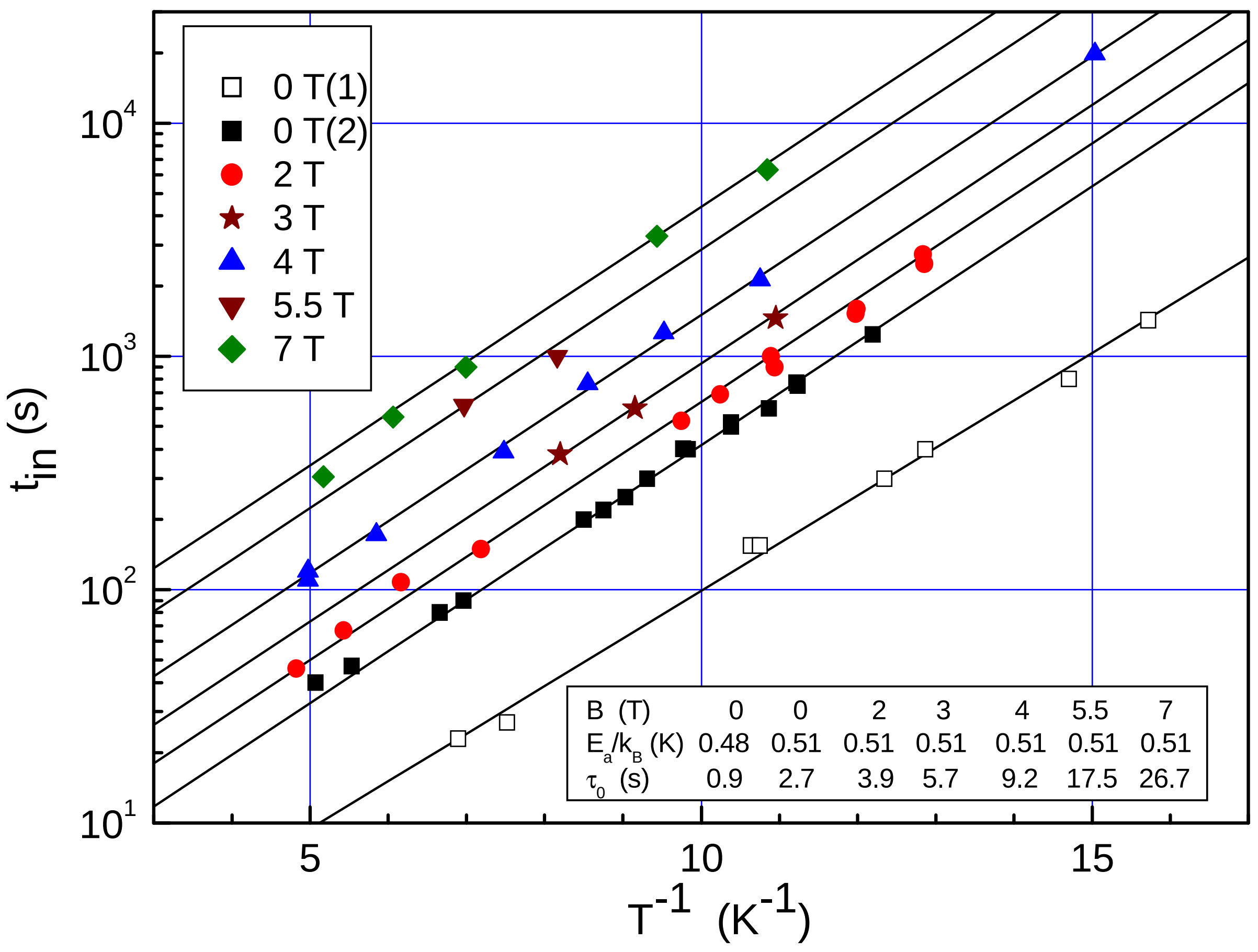

**Fig. 7:** Arrhenius plot of the end of the relaxation time of low energy excitations, $t_{in}$ ,vs. 1/T. The straight lines correspond to an activation energy $E_a/k_B$ = (0.50 ± 0.02) K, almost independent of the field (for the exact fit parameters $E_a$ and $\tau_0$ : see the inset in the Fig.).

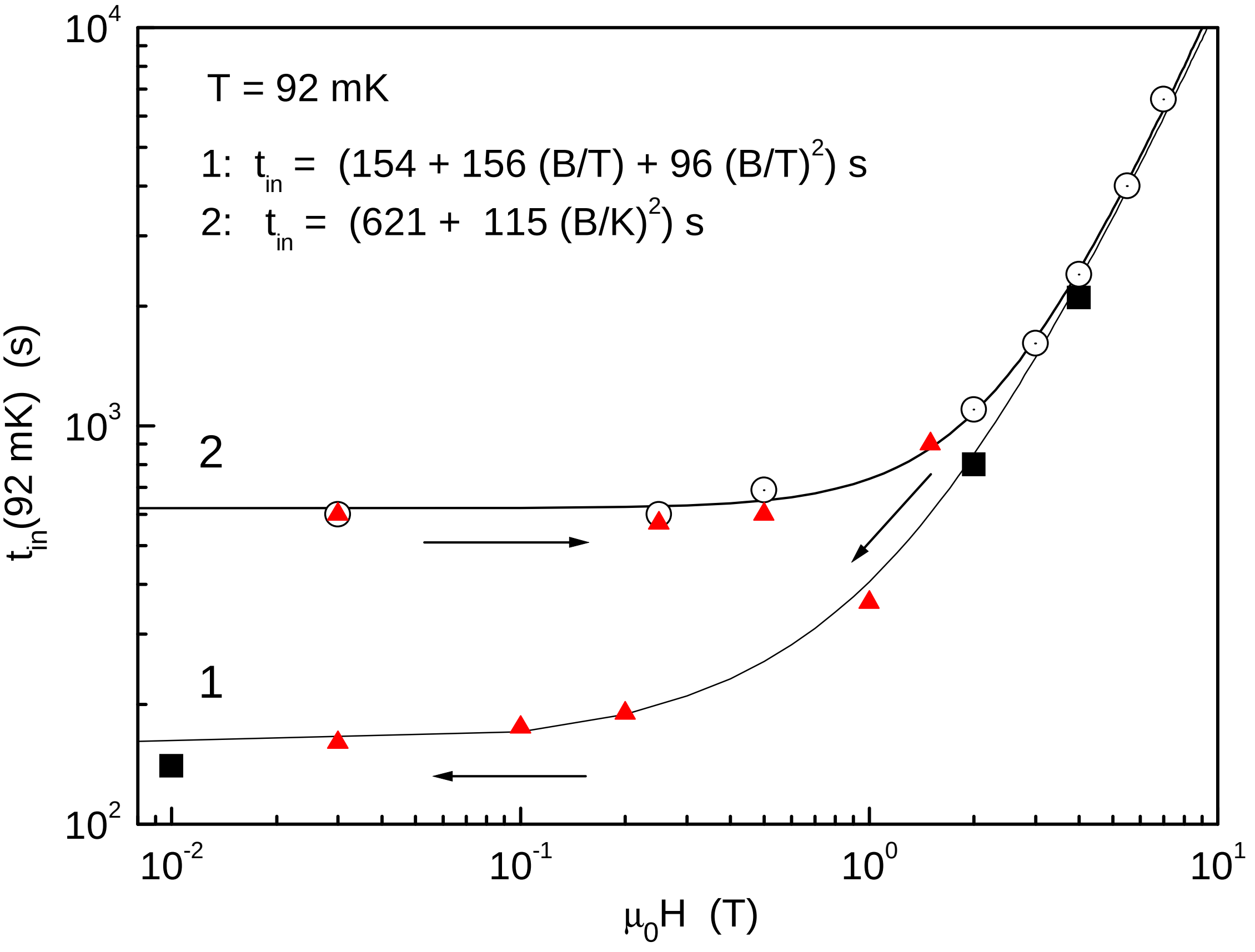

**Fig. 8:** $t_{in}$ (final relaxation time) vs. magnetic field at T = 92 mK. Variations of $t_{in}$ are very similar to those of $c_{eq}$ (see Fig, 7), with the two metastable branches at low field, and the quadratic $B^2$ above 2*T* . Lines are the fits by the expressions given in the inset. For the branch 2, one recovers the same relation as for $c_{eq}$ : sum of a constant plus a $B^2$ term.

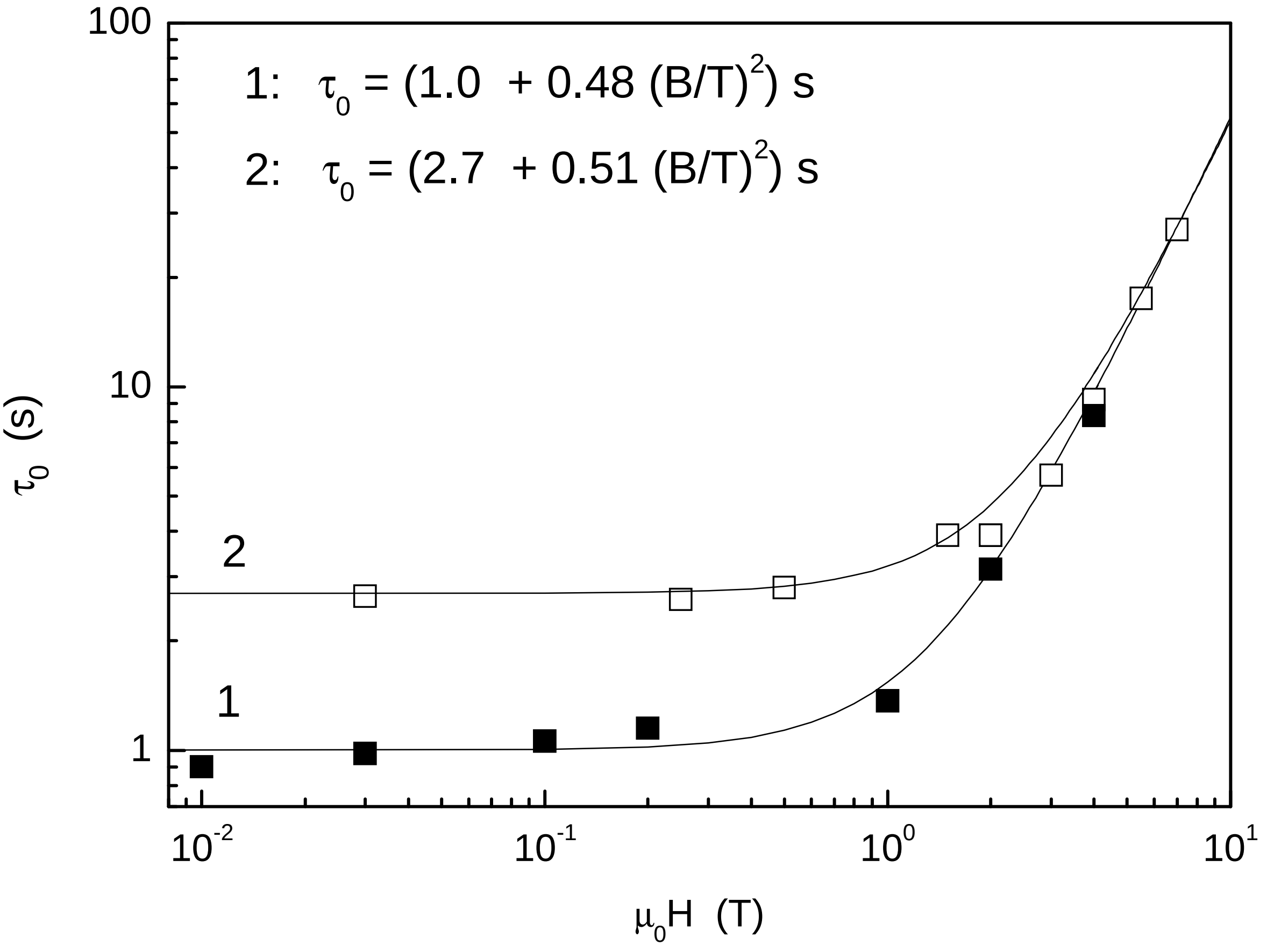

**Fig. 9:** Variations of $\tau_0$, related to the thermally activated regime, vs. magnetic field. The field dependence of $t_{in}$ is mainly caused by the field dependence of $\tau_0$, while the activation energy is nearly field independent.

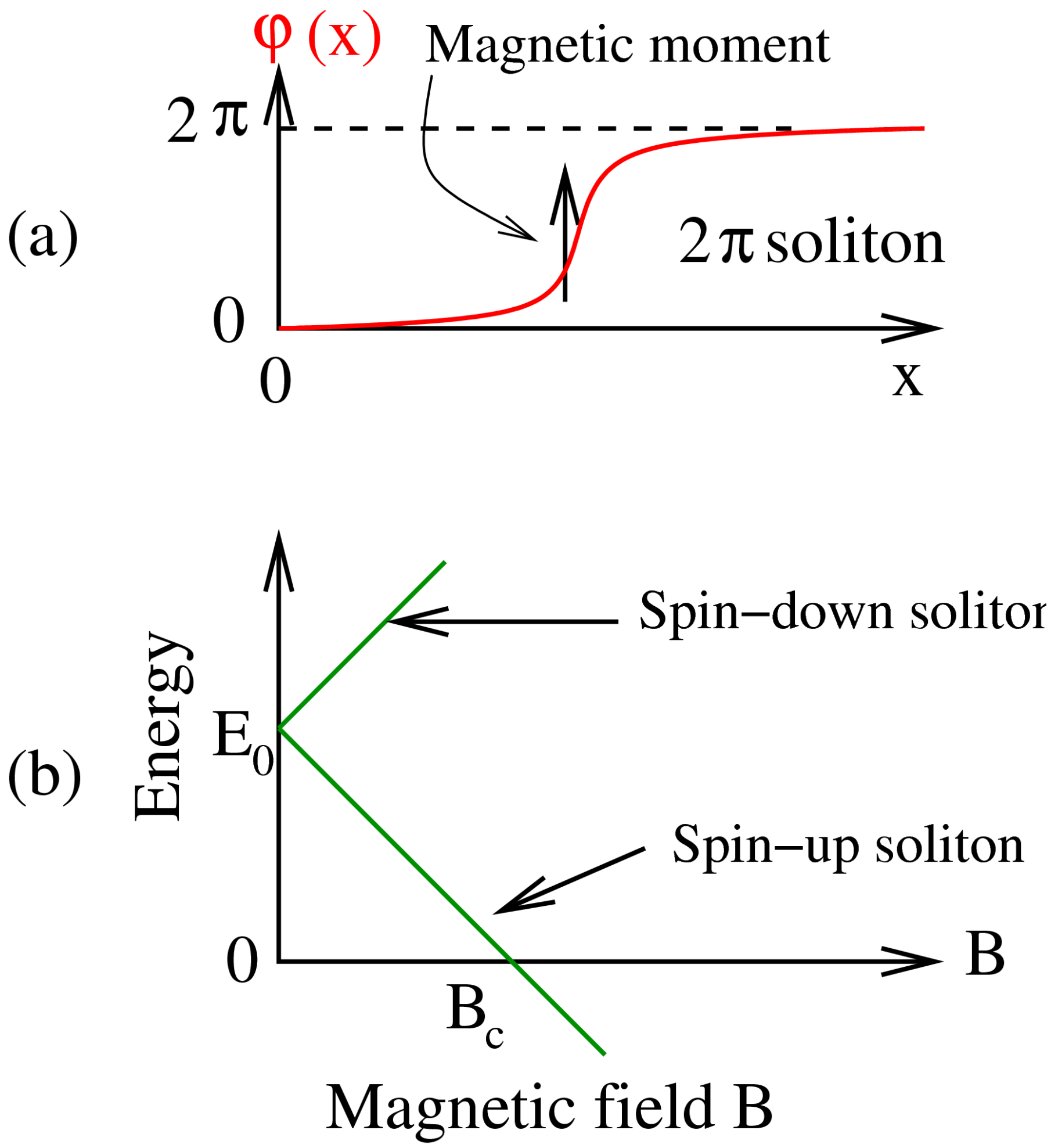

**Fig. 10:** (a) Schematic representation of a soliton in a spin density wave (SDW) carrying a spin-up or spin-down moment. (b) Zeeman splitting of a soliton in a magnetic field. The state with a soliton aligned with the magnetic field is energetically favourable if the magnetic field B exceeds a threshold field $B_c$ .

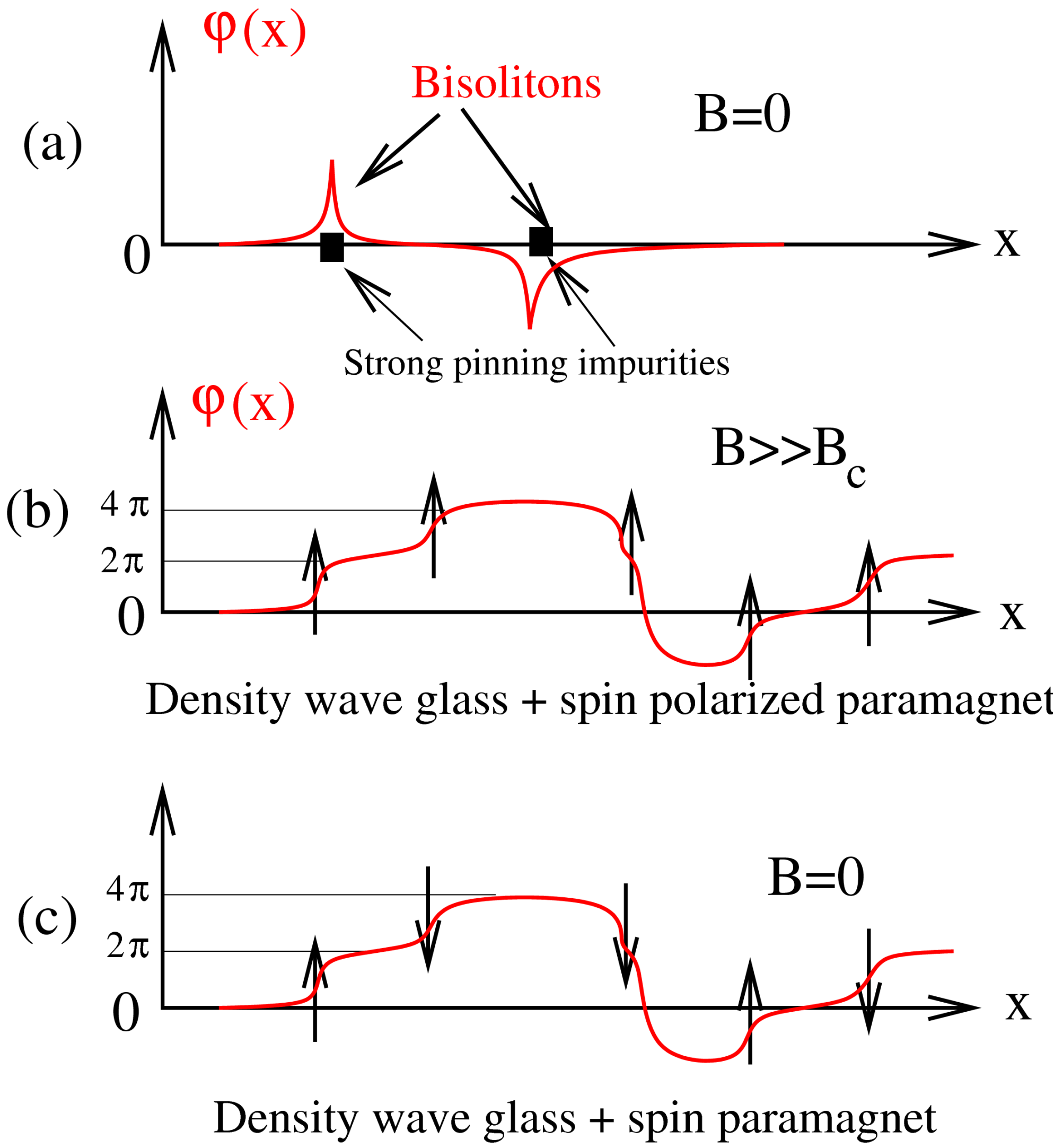

**Fig. 11:** Schematic representation of the Larkin-Ovhinnikov bisolitons in zero magnetic field (a) located at the impurity sites. (b) shows the random phase configuration carrying spin polarized magnetic moments in large field. (c) shows the remanent phase configuration in zero field, carrying magnetic moments with a random orientation.